\documentclass[11pt,a4paper]{article}
\usepackage[T1]{fontenc}
\usepackage[utf8]{inputenc}
\usepackage{lmodern}
\usepackage[margin=1in]{geometry}
\usepackage{amsmath,amsfonts,amssymb}
\usepackage{array}
\usepackage{graphicx}
\usepackage{textcomp,verbatim,multirow}
\usepackage[labelfont=bf]{caption}
\usepackage{subcaption}
\usepackage[round,authoryear]{natbib}
\usepackage{authblk}
\usepackage{adjustbox}
\usepackage{microtype}
\usepackage[hidelinks]{hyperref}

\date{}

\title{Automated carotid artery atherosclerotic inflammation segmentation in PET using MSR-Net}
\author[1,2]{Ran Li}
\author[1,2]{Abhinav K. Jha}
\author[1,2]{Jie Zheng}
\author[1]{Richard Laforest}
\author[1]{Yongjian Liu}
\author[1,3]{Mohamed Zayed}
\author[1]{Tyler J. Fraum}
\author[1]{Barry A. Siegel}
\author[1]{Malak Itani}
\author[1]{Joyce C. Mhlanga}
\author[1,2]{Pamela K. Woodard\thanks{Corresponding author: \href{mailto:woodardp@wustl.edu}{woodardp@wustl.edu}.}}
\affil[1]{Mallinckrodt Institute of Radiology, Washington University in St. Louis, St. Louis, the United States}
\affil[2]{Department of Biomedical Engineering, Washington University in St. Louis, St. Louis, the United States}
\affil[3]{Department of Surgery, Washington University in St. Louis, St. Louis, the United States}
\begin{document}
\maketitle
\begin{center}\footnotesize
\textbf{Author email addresses:} \nolinkurl{ran.li@wustl.edu}, \nolinkurl{a.jha@wustl.edu}, \nolinkurl{zhengj@wustl.edu}, \nolinkurl{rlaforest@wustl.edu}, \nolinkurl{yongjianliu@wustl.edu}, \nolinkurl{zayedm@wustl.edu}, \nolinkurl{tylerjfraum@wustl.edu}, \nolinkurl{siegelb@wustl.edu}, \nolinkurl{mitani@wustl.edu}, \nolinkurl{jmhlanga@wustl.edu}, \nolinkurl{woodardp@wustl.edu}
\end{center}

\begin{abstract}

\noindent\textit{Objective.} To develop an automated method for segmentation and uptake quantification of carotid atherosclerotic inflammation on $^{64}$Cu-CANF-comb PET while accounting for partial-volume and tissue-fraction effects.

\noindent\textit{Approach.} Synthetic carotid PET images with known fractional-volume ground truth were generated by combining MR-derived high-resolution activity maps with GATE Monte Carlo simulation incorporating major PET physical processes and matched to the Biograph mMR acquisition and reconstruction. A two-stage network first localized the carotid vessel wall from T1-weighted MR and then estimated voxel-wise inflammation fraction from masked PET. Both stages used a multiscale residual (MSR) convolutional backbone. Performance was compared with U-Net, Transformer U-Net, DUCK-Net, and iterative Yang partial-volume correction.

\noindent\textit{Main results.} In a two-alternative forced-choice study, five nuclear-medicine physicians identified real images with 54\% accuracy (54/100; Wilson 95\% CI 0.44--0.63; exact binomial $p=0.48$). On 184 simulated test slices from five held-out patients, the proposed method achieved a weighted DSC of 0.79, compared with 0.73 for U-Net, 0.69 for Transformer U-Net, and 0.75 for DUCK-Net, with corresponding HD values of 1.85, 2.29, 2.32, and 2.23 voxels. Patient-cluster bootstrap comparisons favored the proposed method for DSC and HD against all three baselines (Holm-corrected $p\leq0.031$). Overall ENRMSE was 0.24 versus 0.59, 0.68, and 0.59 for the three DL baselines and 0.57 for iterative Yang. In seven patients, corrected SUV showed a positive association with MR-derived lipid-rich/necrotic core with hemorrhage.

\noindent\textit{Significance.} The proposed framework improved carotid PET segmentation and uptake quantification under simulated conditions. The patient analysis is preliminary and requires external validation with biological reference standards before clinical deployment.

\end{abstract}

\noindent\textbf{Keywords:} Atherosclerosis, inflammation, segmentation, PET, deep learning.

\section{Introduction}
Atherosclerosis is the underlying cause of myocardial infarction and ischemic stroke, and a major contributor to coronary artery disease, the leading cause of death worldwide \citep{Intro_WHO}. Recent studies indicate that the inflammation of carotid atherosclerosis contributes to plaque development and rupture, thrombo-occlusion, and cerebral ischemia and stroke \citep{Intro_carotid1} \citep{Intro_carotid2} \citep{Intro_carotid3}. Noninvasive imaging methods such as computed tomography (CT) and magnetic resonance imaging (MRI) can be utilized to assess stenosis caused by carotid plaque and provide information on plaque composition and vulnerability. However, these approaches are unable to detect molecular information related to inflammation. Advanced noninvasive imaging methods such as $^{18}$F-fluorodeoxyglucose positron emission tomography (FDG PET) have been used for identification of inflammation of atherosclerotic plaque \textit{in vivo} \citep{Intro_PET1}. However, FDG PET is nonspecific to inflammation and shows signal in any area of increased metabolic activity. Recently, PET imaging with $^{64}$Cu C-type atrial natriuretic factor comb nanoparticles ($^{64}$Cu-CANF-comb) has been suggested to be a more specific biomarker of inflammation when compared to PET imaging with FDG tracer \citep{Intro_PET2}. Radiotracers that localize on specific targets of inflammation hold promise for more precise identification of inflammation within atherosclerotic plaque and purport to be better predictors of patient outcome. 

Numerous studies have investigated inflammation within carotid plaques \citep{Intro_inf1,Intro_inf2,Intro_inf3,Intro_inf4,Intro_inf5}. In those studies, physicians typically visually analyze PET images alongside anatomical data from conventional imaging modalities such as CT or MRI. The standardized uptake value (SUV) plays an essential role in determining the presence of the inflammatory biomarker being targeted \citep{Intro_SUV1}. Elevated SUV regions within the arterial vessel wall suggest potential inflammatory carotid plaque. However, accurately localizing focal regions of inflammation by delineating contours around visible regions of radiotracer uptake to obtain a SUV is challenging due to the partial volume effects (PVEs) caused by spatial resolution limitations \citep{Intro_SUV2}. This includes spillover effects, where signals from high uptake regions, such as blood pool, extend into adjacent lower uptake regions within the vessel wall.   In addition, spill-out effects complicate manual localization of focal inflammation uptake signals and accurate SUV evaluation in PET images. Various methods, including geometric transfer matrix  \citep{Intro_SUV_cor1}, region-based voxel-wise correction \citep{Intro_SUV_cor2}, iterative Yang \citep{Intro_SUV_cor3}, and reblurred Van-Cittert \citep{Intro_SUV_cor4} have been developed to correct PVEs and spillover effects to some extent. However, these methods have several drawbacks, including reliance on substantial user input---such as manual region of interest (ROI) initialization and contour refinement---high sensitivity to parameter settings, where small changes in thresholds or regularization parameters can lead to large variations in segmentation boundaries and uptake measurements, and dependence on well-aligned MR or CT images when available, collectively limiting their accessibility, robustness, and accuracy. 

In recent years, deep learning (DL) has shown great potential in medical image segmentation. With advancements in DL techniques, many methods have emerged to automatically segment lesions on PET images \citep{Intro_PET_seg1,Intro_PET_seg2,Intro_PET_seg3,Intro_PET_seg4,Intro_PET_seg5,Intro_PET_seg6}. However, all these methods rely on generating manual delineation-derived ground truth, which is time-consuming and subject to intra- and inter-reader variability, especially under the influence of PVEs\citep{Intro_manual_seg1}. To address this issue, strategies have been proposed to use simulations to generate images with known ground truth boundaries, and then use these methods for training. In the context of PET, \citet{Intro_manual_seg7_1} developed such a simulation-based strategy and demonstrated its effectiveness in both reducing the number of patients required for training the network, as well as being less sensitive to PVEs. While this method accounted for the limited spatial resolution of PET systems, the method had limitations in modeling the tissue-fraction effects in PET, namely that a specific voxel in PET can belong to more than one region. To address this issue, \citet{Intro_PET_seg7,Intro_PET_seg8} recently introduced a Bayesian method that estimates the fractional volume occupied by tissue within each voxel of a PET image. This method differs from the conventional method of classifying each voxel in a PET image to a specific region. The promising results provided by that approach encourage advancing that method for segmenting regions of inflammation. Simulation-based training may introduce domain shift because simulated and real PET images can differ in scanner response, reconstruction, noise, and tracer distribution. Building on these advances in PET segmentation research \citep{Intro_PET_seg9,Intro_PET_seg10,Intro_PET_seg11,Intro_PET_seg12,Intro_PET_seg13}, the present study develops an automated approach for carotid plaque PET that estimates voxel-wise fractional inflammatory occupancy within the vessel wall without manual PET region initialization. 

Most previous PET segmentation methods were developed for relatively compact lesions. Carotid PET is different because the arterial wall is thin and lies next to a high-activity blood pool. Previous vascular PET studies have therefore focused mainly on partial-volume correction using anatomically defined regions, including MR-guided and region-based methods \citep{Vasc_Rudd2008,Vasc_Izquierdo2009,Vasc_Bini2014,Vasc_Burg2013}. These methods improve uptake quantification but do not provide an automated voxel-wise estimate of fractional inflammatory occupancy.

Building on this motivation, we propose an end-to-end DL framework specifically designed for carotid plaque uptake segmentation and quantification on PET. The first contribution of our work is that instead of relying on manual ground truth, which is impractical due to the intrinsic low resolution of PET, we develop a comprehensive simulation pipeline to generate realistic carotid PET datasets. The second contribution is that to accurately model partial volume effects (PVEs) and tissue-fraction interactions, we incorporate the method of \citet{Intro_PET_seg8} to estimate the fractional volume of inflammation uptake regions and integrate PET simulation using the GATE Monte Carlo toolkit \citep{Intro_GATE} to model major PET physical processes and increase physical realism. Finally, we also introduce an enhanced two-stage neural network based on \citet{Intro_two_stage}, augmented with a custom multiscale residual (MSR) convolutional backbone for improved feature extraction and robustness.
Together, these innovations are intended to reduce dependence on manual PET delineation, improve uptake quantification in small structures affected by PVEs, and provide an automated workflow for carotid PET analysis. The method is evaluated primarily under simulated conditions and illustrated in a preliminary patient analysis.

\section{Proposed Method}

\subsection{PET/MR imaging protocol}
Patients underwent simultaneous PET and MRI examination of the carotid arteries on a PET/MRI system (Biograph mMR, Siemens Healthineers, Erlangen, Germany) equipped with a bilateral 4 channel phased-array carotid coil (Siemens) for MRI signal reception. The coil allows for sub-millimeter ($\sim$0.5 mm) image resolution to visualize carotid lumen and vessel wall anatomy, while PET can provide physiological information with an in-plane resolution about 4 $mm$. After the MRI scout imaging to localize the carotid arteries in 3 orthogonal planes, a 2-point (dual echo) Dixon scan for attenuation correction (AC) was performed. The PET imaging was then acquired in a list-mode over a 30-min scan, during which MR acquisitions occurred simultaneously. To differentiate various plaque components (lipid rich necrotic core with and without hemorrhage, fibrotic tissue, and calcification), multi-contrast imaging was performed \citep{Method_imging1,Method_imging2,Method_imging3}. T1-weighted MR images were used for vessel-wall segmentation in the proposed network, whereas T2-weighted and PD images were used only for the plaque-component analysis in the patient study. Specifically, Dark Blood turbo spin-echo (TSE) sequence was used to acquire T1-weighted, T2-weighted, and proton-density (PD) images without ECG triggering. Additionally, bright blood T1-weighted 3D gradient-echo sequence was also performed. Superior and inferior
saturation bands were used to suppress blood signal. Imaging parameters for T1-weighted images were: TR/TE=700 $ms$/8 $
ms$. FOV=140$\times$ 140 $mm^2$, and
matrix size=256 $\times$ 256 was interpolated to 512 $\times$ 512, pixel size=0.27 $\times$ 0.27 $mm^2$, bandwidth=345
$Hz/pixel$, echo train length=13, signal average= 2, 16 slices, no slice gap, slice thickness=3 $mm$. For T2-weighted and proton-density imaging, TR/TE = 2000 $ms$/45 $
ms$  and 2000 $ms$/8 $
ms$, respectively.  In the case of bright blood T1-weighted 3D gradient-echo imaging, the parameters were TR/TE = 25 $ms$/3.5 $
ms$.

The 30-min PET data were used 
for high-definition (HD) PET imaging reconstruction, performed  
using a point-spread-function-corrected ordinary Poisson ordered subsets expectation maximization
(PSF-OP--OSEM, or simply PSF-OSEM) algorithm with optimal iterations and subsets, and a Gaussian smoothing filter. The size of the image matrices was 
172\ensuremath{\times}172\ensuremath{\times}127, resulting in a pixel size of 4.17 $mm$ and a slice thickness of 2.03 $mm$. MR-based 2-
point Dixon AC for the PET image was performed to calculate radiotracer uptake.

\subsection{Synthetic PET data generation}
Here we present a data simulation pipeline designed to address the challenge of the absence of ground truth data for training DL models. First, PET images were resampled to match the slice direction, field of view, and matrix size of the corresponding T1-weighted MR images using MIM (MIM Software Inc., version 7.1.5). Residual alignment was manually refined using the carotid bifurcation and jugular vein as landmarks. Only in-plane translation and through-plane slice position were adjusted; rotation and scaling were not applied. Inter-operator variability of this manual refinement was not prospectively measured. To identify potential regions of inflammation within the carotid walls, we delineated the contours of the carotid walls on T1-weighted MR images.

These MRI-aligned high-resolution PET images were used to generate high-resolution background radioactivity maps. To avoid the potential influence of regions of inflammation within carotid walls, radioactivity within the vessel-wall region was constrained to be under 1500 $Bq/mL$. This value was selected from 48 manually labeled candidate uptake regions in 23 patients, for which the minimum, 5th percentile, and median activity concentrations were 1530, 1610, and 2050 $Bq/mL$, respectively.

To diversify the synthetic dataset and encompass multiple possible PET scanning scenarios, the intensities of aligned high-resolution PET images were min--max normalized to the range of 0 to 1, followed by rescaling to a range of 0 to $\alpha$. The maximal value of $\alpha$ was randomly selected from 2500 to 3500 $Bq/mL$, determined from real patient data post-attenuation correction. This rescaling was used to construct activity maps in absolute units for GATE simulation and was separate from network input normalization. The shape of synthetic regions of inflammation within arterial walls was randomly simulated using closed B\'ezier curves \citep{Method_Bezier}. The minor and major axis lengths were uniformly selected from 4 to 7 $mm$ and 7 to 9 $mm$, respectively, guided by measurements from the patient cohort. Based on statistical analysis of manually delineated inflammation regions in our database, the radioactivity concentration was sampled from 1500 to 2500 $Bq/mL$, corresponding to SUVs of approximately 1.7 to 3 for a 70-kg patient. Lesion centroids and orientations were randomly sampled within the vessel wall, and contours extending beyond the wall were clipped to the wall mask. Placements retaining less than 25\% of the sampled lesion area were resampled. The synthetic and patient distributions showed no detectable differences in major axis ($D=0.14$, $p=0.62$), minor axis ($D=0.12$, $p=0.71$), or activity concentration ($D=0.15$, $p=0.54$).

Following the insertion of inflammatory regions, the absolute radioactivity concentration was computed for each high-resolution voxel ($0.27mm \times 0.27mm \times 3mm$) to generate high-resolution activity maps (HRMs), which served as known high-resolution radioactivity distributions for the forward PET simulation. The HRM intensity image itself was not used as the direct Stage-2 training target. Stage 2 was designed to estimate tissue occupancy within each reconstructed PET voxel rather than to reconstruct the absolute high-resolution activity field. Accordingly, the ground truth was defined as the fractional volume occupied by inflammation within each reconstructed PET voxel. This representation decouples the training target from the absolute lesion activity concentration: lesions with the same geometry but different activity concentrations have the same fractional-occupancy target. It also expresses the target as a dimensionless quantity on the reconstructed PET voxel grid, bounded between 0 and 1.

Let $\omega^{\text{PET}}_m(\mathbf{r})$ and $\omega^{\text{HRM}}_n(\mathbf{r})$ denote binary indicator functions for the $m$-th reconstructed PET voxel and the $n$-th HRM voxel, respectively. Let $\epsilon^{\text{HRM}}_n$ equal 1 when the $n$-th HRM voxel belongs to synthetic inflammation and 0 otherwise. Let $N$ be the number of HRM voxels and $V$ the physical volume of a reconstructed PET voxel. The fractional ground-truth volume of inflammation for the $m$-th PET voxel is defined as \citep{Intro_PET_seg7}:
\begin{equation}
v_{m}  = \frac{1}{V}
\sum_{n=1}^{N} 
\epsilon^{\text{HRM}}_{n}
\int d^3\mathbf{r} \,
\omega^{\text{HRM}}_n(\mathbf{r})\,
\omega^{\text{PET}}_m(\mathbf{r}).
\end{equation}

This formulation computes the normalized volumetric overlap between high-resolution inflammation voxels and each reconstructed PET voxel, yielding a continuous fractional label between 0 and 1. The fractional target isolates tissue occupancy from the absolute activity scale, while the reconstructed PET intensity retains the uptake information used for downstream quantification. By defining lesion boundaries at high spatial resolution and subsequently simulating PET acquisition and reconstruction, this framework models both sources of PVEs: those arising from the limited spatial resolution of the PET system and those due to tissue-fraction effects within individual PET voxels. Direct prediction of the HRM would instead require simultaneous recovery of high-resolution geometry and absolute radioactivity concentration from lower-resolution reconstructed PET, constituting a different super-resolution/inverse-reconstruction task.

A virtual model of the Siemens Biograph mMR used for the patient studies was implemented in GATE 8.2 with Geant4 10.5.1. The Biograph mMR has a reported NEMA transverse spatial resolution of approximately 4.4 mm FWHM at 1 cm from the center of the field of view \citep{Method_mMR}. The major scanner and detector settings are summarized in Table~\ref{tab:gate_params}.

\begin{table}[htbp]
\centering
\begin{adjustbox}{max width=\linewidth}
\begin{tabular}{ll}
\hline
Parameter & Value \\
\hline
Crystal material & LSO \\
Crystal dimensions & $4 \times 4 \times 20$ mm \\
Block configuration & $8 \times 8$ crystals \\
Blocks per ring & 56 \\
Block rings & 8 \\
Crystal rings & 64 \\
Detector ring diameter & 656 mm \\
Axial field of view & 258 mm \\
Crystals per ring & 448 \\
Energy resolution at 511 keV & 11.5\% \\
Energy acceptance window & 430--610 keV \\
Coincidence timing window & 5.86 ns \\
Delayed-window offset (randoms) & 500 ns \\
Multiple-coincidence policy & takeWinnerIfAllAreGoods \\
Dead-time model & Not modelled \\
Readout level & Block \\
\hline
\end{tabular}
\end{adjustbox}
\caption{Geometry and detector-response settings of the Biograph mMR model implemented in GATE. The published mMR characterization is cited for system context \citep{Method_mMR}.}
\label{tab:gate_params}
\end{table}

\begin{figure}[!t]
\centering
\includegraphics[
    width=0.85\textwidth,
    height=0.35\textheight,
    keepaspectratio
]{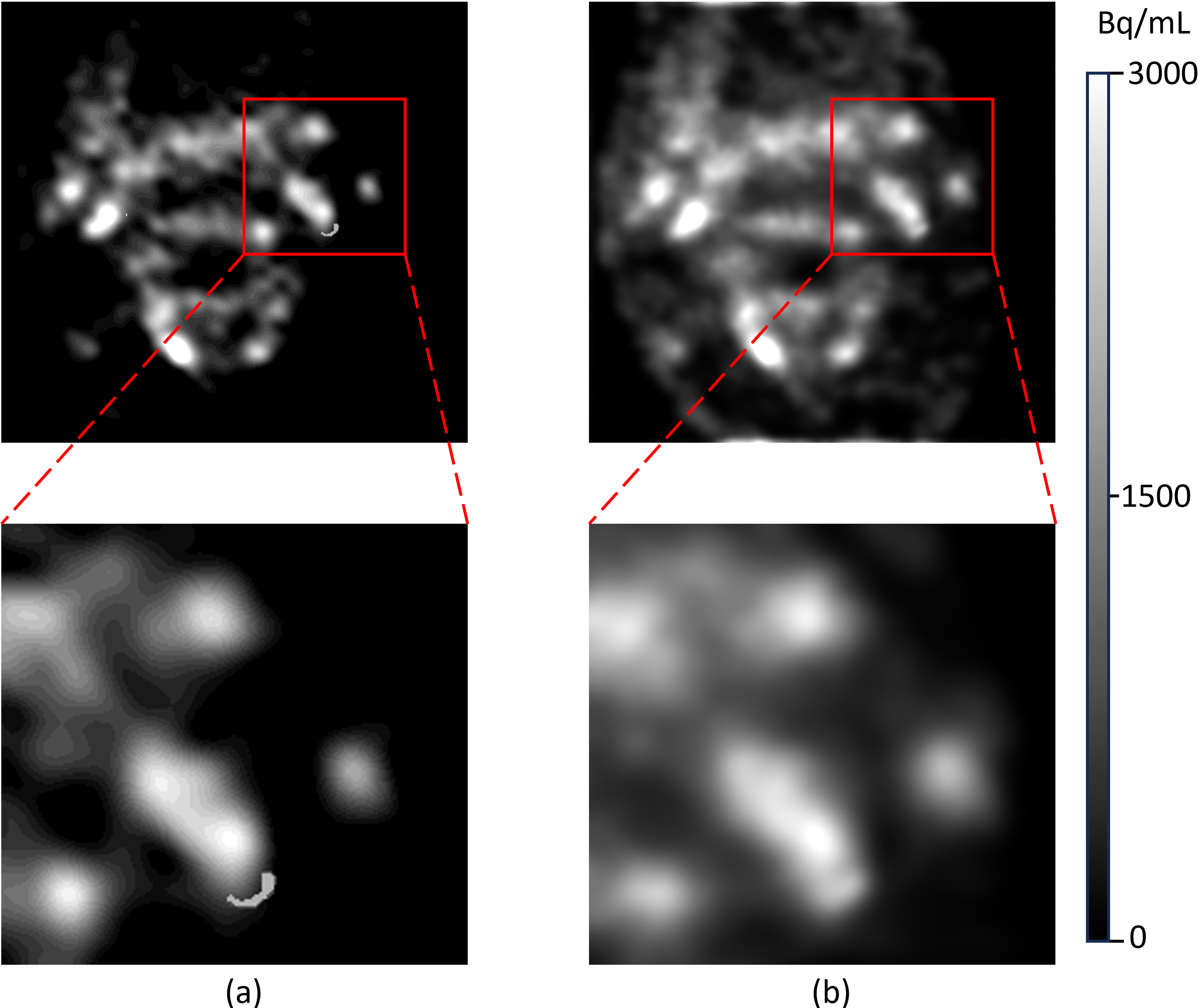}
\caption{Representative simulation example: (a) high-resolution uptake map with a simulated high-uptake region within the vessel wall (left column) and (b) the corresponding synthetic PET image (right column).}
\label{fig:syntic}
\end{figure}

The HRMs were used as 3D voxelized $^{64}$Cu sources, and a 30-minute list-mode acquisition was simulated. The simulation included the $^{64}$Cu positron spectrum and transport, annihilation non-collinearity, photon attenuation and scatter, and detector interactions. Coincidences were processed using a 430--610 keV energy window and a 5.86-ns timing window. Images were reconstructed in CASToR \citep{Method_Castor} using PSF-OP-OSEM with 21 subsets and 8 iterations followed by 4.0 mm Gaussian filtering, with attenuation, scatter, randoms, and normalization corrections. The reconstructed matrix was $34\times34$ with $4.17\times4.17\times2.03$ mm voxels and was interpolated to $512\times512$ to match the MR images. A representative high-resolution uptake map and the corresponding reconstructed synthetic PET image are shown in Fig.~\ref{fig:syntic}.

\subsection{Two-stage neural network with novel convolution backbone}
\label{sec:network}

\begin{figure}[htbp]
\centering
\includegraphics[width=\linewidth]{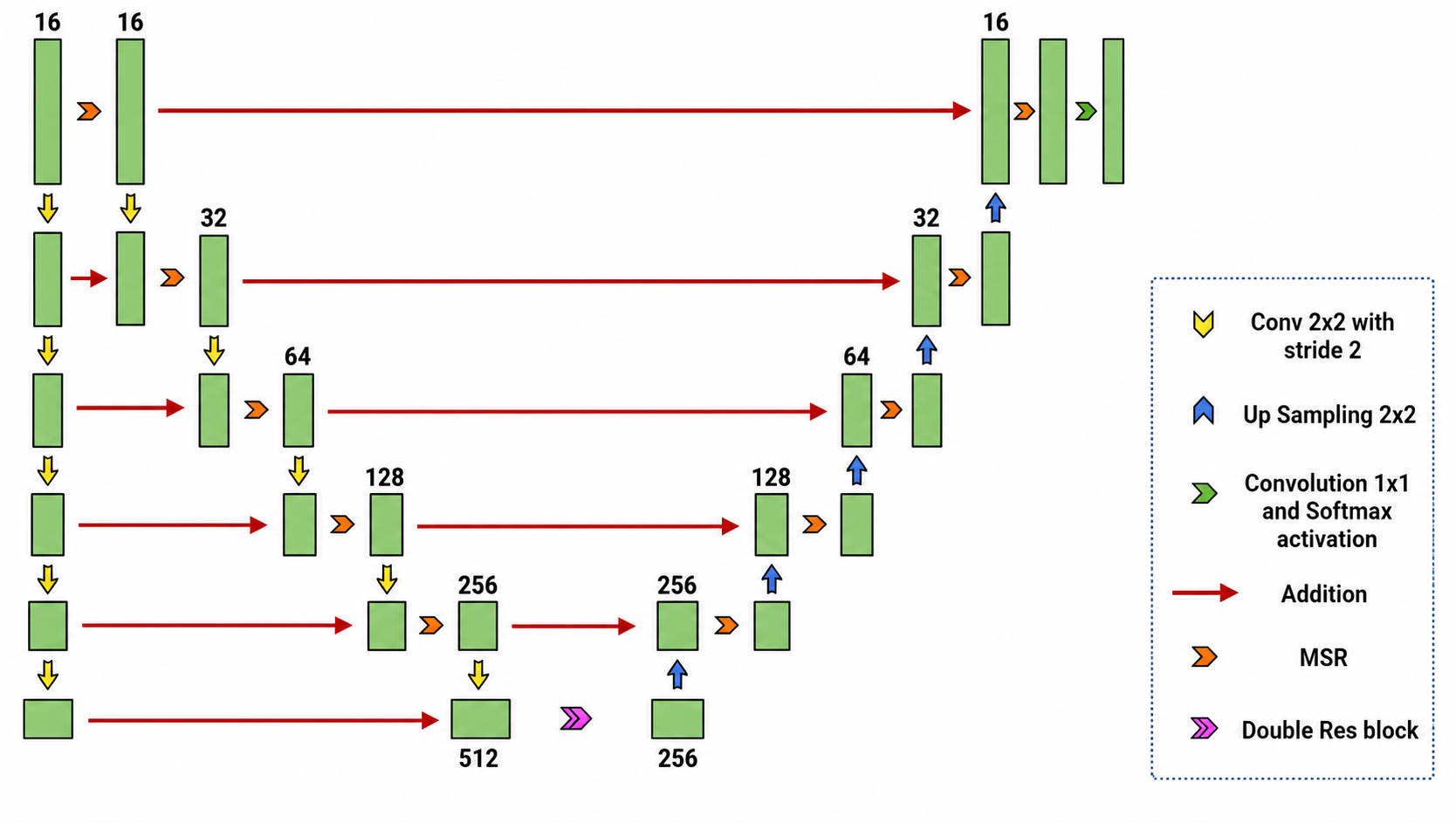}
\caption{Structure of neural network.}
\label{fig_network}
\end{figure}
The two stages use the same network architecture but different inputs and targets. Stage 1 receives a T1-weighted MR image and predicts a binary vessel-wall mask. The predicted mask is morphologically enlarged and applied to the PET image by element-wise multiplication. Stage 2 receives the masked PET image and estimates the continuous fractional inflammation volume defined in Equation~(1). The two stages are trained sequentially; masks generated by the trained Stage-1 network are used for Stage-2 training and inference. This design separates the MR-based anatomical localization task from the simulation-based fractional-volume estimation task.

The neural network architecture, as shown in Fig. \ref{fig_network}, is based on UNet \citep{Method_unet}, comprising 6 encoder blocks with multiscale input and a decoder. The additional $2\times2$ stride-2 convolutions downsample the multiscale image input to the spatial resolution of each encoder level before fusion. Learned convolutions were used instead of max pooling to preserve weak distributed activity while allowing the downsampling operation to be optimized during training.

Our proposed segmentation neural network incorporates two key innovations aimed at enhancing performance in identifying regions of inflammation within carotid walls. First, we introduce an additional neural network component with the same structure for segmentation of the arterial vessel wall. To account for spillover and residual PET--MR misalignment, the output vessel-wall mask was morphologically enlarged to approximately four times its original size. Morphological enlargement factors of $1\times$, $2\times$, $3\times$, $4\times$, $6\times$, and $8\times$ relative to the original mask were evaluated, with $4\times$ providing the best overall segmentation and uptake-quantification performance.

\begin{figure}[htbp]
\centering
\includegraphics[width=\linewidth]{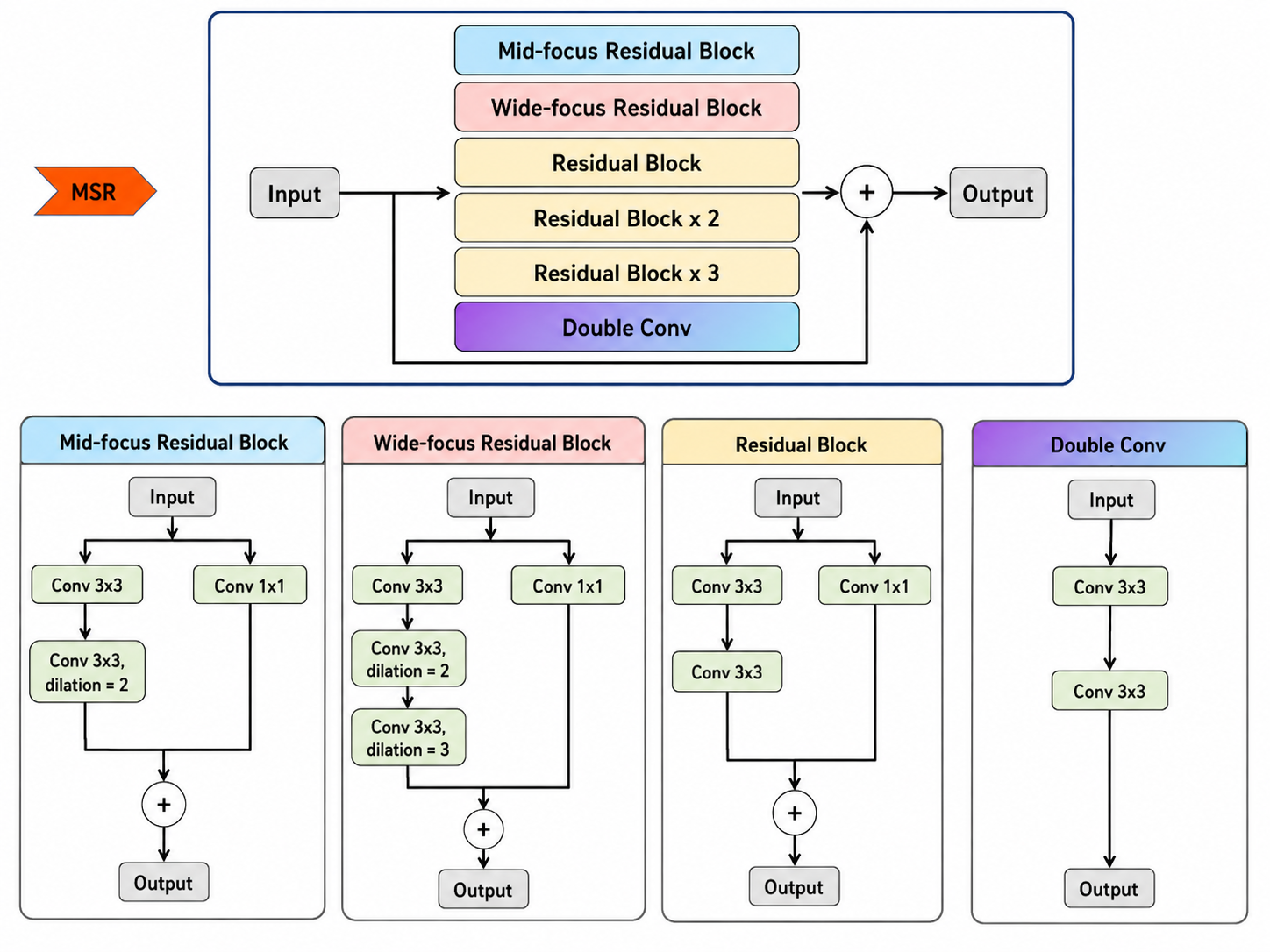}
\caption{(a) Structure of MSR; (b) Structure of Mid-focus Residual Block; (c) Structure of Wide-focus Residual Block; (d) Structure of Residual Block; (e) Structure of double convolution.}
\label{fig_network_part}
\end{figure}

Secondly, inspired by DUCK-NET \citep{Method_DUCK}, we designed a multiscale residual (MSR) block for feature extraction. The MSR block contains wide- and mid-scope dilated residual branches, three standard residual branches, and a double-convolution branch operating in parallel. The three standard residual branches contain one, two, and three consecutive residual blocks, respectively. Because each residual block contains two $3\times3$ convolutions, increasing the number of consecutive residual blocks progressively enlarges the effective receptive field while preserving the spatial resolution of the feature maps. The corresponding effective receptive fields of these three standard residual branches are approximately 5, 9, and 13 pixels, respectively. Thus, the shallower branch emphasizes local features, whereas the deeper branches incorporate progressively broader spatial context before fusion. Batch normalization before branch fusion places the double-convolution output on a comparable scale to the residual branches, and the branch ablation in Section~\ref{sec:ablation} evaluates its contribution. As illustrated in Fig. \ref{fig_network}, the network also integrates multiscale PET inputs at different encoder depths. Fig. \ref{fig_network_part} provides a detailed illustration of the MSR components.

\subsection{Rationale for the multiscale residual backbone}
\label{sec:theory}

The MSR block combines branches with different receptive fields so that local focal uptake and broader spillover/background patterns can be represented in parallel. Addition-based fusion was selected because it combines these branches without increasing the channel dimension and therefore limits parameter growth. The branch-composition and addition-versus-concatenation experiments in Section~\ref{sec:ablation} provide the empirical evaluation of these design choices. The supporting receptive-field and approximation analysis is provided in Appendix~A.

\section{Experimental Details}
\subsection{Data generation and pre-processing}
A total of 45 patients were scanned with a Siemens PET/MR scanner according to our imaging protocol, resulting in the acquisition of 16 MR slices for each patient. The contours of vessel walls were manually delineated in a customized Matlab toolkit to generate ground truth data for vessel wall segmentation on T1-weighted MR images. PET data from 23 patients with good image quality were selected to produce high-resolution radioactivity maps and corresponding ground truth fractional volume of inflammation following our PET simulation pipeline. For each patient, the simulation process was repeated five times, varying the assignment of inflammation and background uptake to generate additional data. To accelerate the PET simulation, we employed 60 computing cores on a high-performance computing cluster, each equipped with a Xeon Gold CPU and 64 GB of memory. Following the simulation pipeline, a total of 1184 reconstructed PET slices were generated. Subsequently, masks generated from T1-weighted MR images were applied to synthetic PET images to focus the attention of the neural network on the regions of interest.

\subsection{Evaluation of the realism of the synthetic PET images}

To assess the realism of the synthetic PET images, we conducted a Two-Alternative Forced Choice (2AFC) study following the methodology outlined by \citet{Evaluation_PET_realism}. The study involved experienced nuclear-medicine (NM) physicians who possess substantial expertise in interpreting medical images. These participants were selected for their ability to observe subtle differences in images. Each participating NM physician was presented with 20 pairs of images through a custom-designed web application. Each pair consisted of one real patient PET image and one synthetic PET image generated using the simulation pipeline developed in the study. A total of five experienced NM physicians took part in the study. Observers were tasked with identifying which image in each pair they believed to be the real PET image and rating their confidence in each decision on a five-point scale (5 = high confidence, 1 = low confidence). This forced-choice format ensured that observers made direct comparisons between the two images in each pair. The accuracy of each observer's responses was recorded, with accuracy close to 50\% interpreted as indicating that readers could not reliably discriminate synthetic from real PET images under the tested conditions.

\subsection{Network training}
The neural network was trained in two separate stages. From a subset of 23 patients with good image quality, real MR data and corresponding synthetic PET data generated via the proposed simulation pipeline were used for model development. Specifically, data from 18 patients were used for training, while data from the remaining 5 patients were reserved for testing and performance evaluation.

The first stage involves a vessel wall segmentation network aimed at localizing artery regions. To train this stage, 250 T1-weighted MR images from 18 patients, along with corresponding ground truth data generated by manually delineating carotid arteries, were augmented to 1000 using geometric transformations comprising scaling, rotation, translation and flipping. Scaling was sampled uniformly from 0.90 to 1.10, rotation from $-10^{\circ}$ to $+10^{\circ}$, translation from $-10$ to $+10$ pixels in each in-plane direction, and left--right reflection was applied with probability 0.5. The 10\% scaling limit was selected to perturb vessel dimensions without generating lumen or wall sizes outside the range observed in the source cohort. The hyperparameters of the neural network were optimized via five-fold cross-validation on the training dataset. Additionally, 46 T1-weighted MR images from an additional 5 patients were allocated for testing.

In the second stage, the focus was on segmenting and quantifying the tissue fraction of regions of inflammation on PET images. For this stage, the 1000 synthetic PET slices from the 18 development patients were augmented and used for training, while 184 PET slices from the five held-out patients were used for evaluating model performance. As in the first stage, the hyperparameters of the neural network were optimised through five-fold cross-validation on the training dataset. Cross-validation was performed at the patient level using only the 18 development patients and was used for hyperparameter selection. All slices from one patient remained in the same fold. After hyperparameters were selected, a single model was retrained using all 18 development patients and evaluated once on the five held-out test patients; no test-patient data entered model selection. For both stages, dropout was incorporated into the network during training to reduce overfitting, and early stopping based on validation performance was used to terminate training when no further improvement was observed.

Both stages of model training employed the cross-entropy loss function, defined as:
\begin{equation}
\mathcal{L}_{ce} = - \left[ y \log(\hat{y}) + (1 - y)\log(1 - \hat{y}) \right].
\end{equation}
In the first stage (vessel wall segmentation), $y$ denotes the binary ground truth label indicating whether a voxel belongs to the vessel wall, and $\hat{y}$ represents the predicted probability.

In the second stage (inflammation estimation), $y$ denotes the ground-truth fractional volume of inflammation within each voxel and $\hat{y}$ denotes the corresponding estimate. The target is a physical tissue-fraction quantity derived from the HRM geometry, not an intensity-normalized HRM value; its bounded range $[0,1]$ follows from the definition of fractional occupancy. The loss formulation follows the tissue-fraction estimation method of \citet{Intro_PET_seg8}, in which segmentation is posed as estimation of a continuous fractional volume rather than assignment of each voxel to a single tissue class. A softmax activation is used at the network output, and $\hat{y}$ denotes the resulting predicted fraction for the inflammation class. Liu \textit{et al.} defined the tissue-fraction estimation cost using binary cross-entropy (BCE) between the true and estimated fractional volumes and showed that minimizing the ensemble BCE cost yields the posterior-mean estimate of the voxel-wise tissue fraction. We therefore use BCE for Stage 2 as an estimation loss for the continuous fractional target, rather than as a conventional binary-classification loss.

The proposed method was trained using high-performance computational resources, specifically a high-performance computing cluster equipped with an NVIDIA Tesla V100 GPU with 32 GB of HBM2 memory.

\subsection{Evaluating the segmentation performance of the proposed method}
To rigorously evaluate the ability of the proposed method to segment inflammation under varying levels of contrast between inflamed regions and the background blood signal, we quantified contrast using the uptake ratio between inflammation and background. In the synthetic PET test dataset, consisting of 184 slices from 5 patients, this contrast metric ranged from 0.6 to 1.8. We divided the test dataset into three contrast intervals: [0.6, 1.0], [1.0, 1.4], and [1.4, 1.8], containing 32, 98, and 54 slices, respectively. This stratification enabled us to assess segmentation performance under different levels of inflammation-to-background contrast that may arise in clinical scenarios. Segmentation performance was then evaluated within each contrast range, focusing on spatial overlap and contour similarity between the predicted segmentation and the ground truth.

We compared the proposed method with several baseline approaches, including conventional U-Net, Transformer U-Net, and DUCK-Net, which are widely used and have demonstrated strong performance in medical image segmentation tasks. All comparison models were trained on synthetic PET images generated using our PET simulation pipeline and optimized within a five-fold cross-validation framework. Each model was trained to classify each voxel into its corresponding region. To ensure a fair comparison, all methods were trained on the same simulated PET dataset. In addition, the architectures of the baseline models, including standard U-Net and Transformer-based U-Net, were configured to have comparable capacity in terms of the number of trainable parameters, to the extent possible given their structural differences. Hyperparameters, training dataset size and data augmentation strategy were kept consistent across all methods. The Transformer U-Net was trained from scratch without ImageNet or other external pretraining. By using identical training data and comparable model capacities, we ensured an unbiased comparison that isolates the contribution of tissue-fraction modeling and the proposed MSR backbone, rather than differences in data exposure, network scale, or training procedure.

\subsection{Figures of merit for evaluation}

\subsubsection{Evaluation using conventional metrics}

The ground truth in our method consists of continuous-valued fractional volumes for each voxel. For calculation of binary segmentation metrics, a voxel was assigned to the inflammation class when its fractional value exceeded 0.5, corresponding to majority occupancy by inflammation. This threshold is applied only for computation of DSC and HD; uptake quantification uses the continuous fractional estimate directly and involves no binarisation. Sensitivity of DSC and HD to the prediction threshold is reported in Table~\ref{tab:threshold_sensitivity}, which shows a broad optimum around 0.5. To quantify spatial overlap between the binarised ground truth and prediction, we employed the Dice similarity coefficient (DSC) \citep{FOM_DSC}, which ranges from 0 to 1, with higher values indicating more accurate segmentation performance. Shape similarity was quantified using the Hausdorff distance (HD) \citep{FOM_HD}, where lower values imply more accurate segmentation. 

\subsubsection{Evaluation on the task of uptake quantification}

Previous studies have shown that segmentation metrics such as the Dice similarity coefficient may not reliably reflect performance in downstream clinical applications, particularly for quantitative tasks \citep{Eval_ENRMSE,Eval_ENRMSE2}. Therefore, it is important to evaluate segmentation methods within the context of clinically relevant objectives. In this study, the clinical task of interest is the estimation of tracer uptake within regions of inflammation.

The total uptake ($U$) within inflammatory regions is defined as the sum, over all voxels, of the product between voxel-wise uptake ($p_n$) and the corresponding fractional volume of inflammation ($f_n$):
\begin{equation}
U = \sum_{n=1}^{N} p_n f_n,
\end{equation}
where $N$ denotes the total number of voxels in the PET image.

To evaluate performance on this task, we assessed the accuracy of estimated inflammation uptake across a dataset of five test patients (184 synthetic PET slices). Consistent with the previous section, evaluations were conducted across three levels of inflammation-to-background contrast. We further compared the proposed method with three baseline architectures: conventional U-Net, Transformer U-Net, and DUCK-Net, to examine the impact of the proposed MSR backbone on uptake quantification performance.

Performance was quantified using the ensemble normalized root mean square error (ENRMSE), defined as:
\begin{equation}
\mathrm{ENRMSE} = \sqrt{\frac{1}{S} \sum_{s=1}^{S} \left( \frac{u_g - u_p}{u_g} \right)^2},
\end{equation}
where $u_p$ and $u_g$ denote the predicted and ground-truth uptake values, respectively, and $S$ is the total number of evaluated samples.

To account for the correlation among multiple slices from the same patient, formal comparisons of DSC and HD between the proposed method and the three deep-learning baselines were performed using a paired patient-cluster bootstrap with 10,000 resamples. The five held-out test patients were sampled with replacement, with all slices from a selected patient retained together as a cluster. For each resample, the paired difference between the proposed method and each comparator was calculated. The 95\% confidence intervals were obtained from the bootstrap distribution, and two-sided $p$ values were Holm-corrected across the three baseline comparisons separately for DSC and HD. The overall ENRMSE was calculated by pooling the squared normalized errors across all test slices before taking the root mean square, rather than by averaging the ENRMSE values of the three contrast subgroups. The same patient-cluster bootstrap framework was used for uptake quantification, with overall ENRMSE recomputed for each resample and the paired ENRMSE difference between the proposed method and each baseline used to obtain 95\% confidence intervals.

\subsection{Quantitative analysis on real patient data}
\label{sec:patient_analysis}

Previous studies have suggested that the degree of plaque inflammation is associated with the extent of lipid-rich or necrotic cores (LR/NC) and intraplaque hemorrhage \citep{Eval_LI_link1,Eval_LI_link2,Eval_LI_link3,Eval_LI_link4}. However, due to limitations in radiotracer specificity and the influence of partial volume effects (PVEs) in PET imaging, a quantitative association between LR/NC with hemorrhage and inflammation has not been well established in patient data.

To evaluate the effectiveness of the proposed method, we conducted a pilot study using $^{64}$Cu-CANF-comb PET/MR data from 7 patients identified by physicians as having a standardized uptake value (SUV) greater than 2.0. PET images were co-registered with the corresponding MR images using MIM software. The MR images were input into the first stage of the neural network to generate vessel wall masks. These masks were then applied to the PET images, which were subsequently processed by the second stage of the network to estimate voxel-wise fractional maps of inflammation.

Unlike conventional approaches that use binary segmentation masks to quantify uptake, the proposed method estimates the fractional contribution of inflammation within each PET voxel. This fractional map was used to correct SUV measurements by accounting for partial volume effects, enabling more accurate quantification of inflammation uptake.

Additionally, carotid plaque-component segmentation, including LR/NC with intraplaque hemorrhage, was generated from the multi-weighted MR images using the previously published two-stage neural-network method of \citet{Intro_two_stage}. These MR-derived plaque-component segmentations were used only for the patient analysis and were not outputs of the PET segmentation network. The analyzed plaques exhibited heterogeneity in volume, composition, and anatomical variability across patients. To account for this heterogeneity, the carotid arteries were divided into four regions---anterior, posterior, lateral, and medial---based on their relative spatial positions. For each slice, we quantified the percentage volume of LR/NC with intraplaque hemorrhage as:
\begin{equation}
P = \frac{V_{L}}{V_{vw}},
\end{equation}
where $P$ denotes the percentage volume, $V_{L}$ is the volume of LR/NC, and $V_{vw}$ is the vessel wall volume.

The percentage volume of LR/NC was then analyzed in conjunction with inflammation uptake derived from PET images. Specifically, corrected SUV was computed using the estimated fractional inflammation map as:
\begin{equation}
\mathrm{SUV}_{\mathrm{corr}} = \frac{M_{IF} \cdot R}{I_n} = M_{IF} \cdot \mathrm{SUV},
\end{equation}
where $M_{IF}$ represents the fractional inflammation map, $R$ is the decay-corrected radioactivity concentration ($\mathrm{Bq/mL}$), and $I_n$ is the injected activity normalized by body weight ($\mathrm{Bq/g}$).

In this framework, atherosclerotic inflammation is quantified using the corrected SUV values derived from the proposed method, while intraplaque hemorrhage is characterized using MR-based LR/NC measurements. The association between LR/NC with intraplaque hemorrhage and corrected SUV was evaluated using linear mixed-effects models with a patient-specific random intercept to account for repeated slice-level observations within each patient. Separate models were fitted for corrected $SUV_{mean}$ and $SUV_{max}$ using 37 observations from seven patients. Fixed-effect slopes are reported per 10-percentage-point increase in LR/NC with 95\% confidence intervals and two-sided $p$ values.

\section{Results}
\subsection{Performance in the realism of the synthetic PET images}
\label{sec:realism_results}

Table \ref{tab:Realism_performance} shows the percent accuracy and median confidence levels for each expert human observer who participated in the 2-AFC study, which aimed to evaluate the realism of synthetic PET images generated by our simulation pipeline. Pooled across the five observers and 20 image pairs each, 54 of 100 forced-choice decisions were correct. This does not differ significantly from chance (exact binomial test against $p_0 = 0.5$, $p = 0.48$; Wilson 95\% confidence interval for the pooled accuracy 0.44 to 0.63).  Considered individually, no reader performed significantly better than chance; the highest observed accuracy, 65\%, corresponds to 13 of 20 correct and is not significant in isolation ($p = 0.26$, uncorrected), and was accompanied by a median confidence of 2. Most observers reported median confidence of 3 or below, indicating general uncertainty in discriminating real from synthetic images.

We note the limited power of this design: with 100 pooled trials, the study can exclude only fairly large departures from chance, and an accuracy as high as approximately 0.63 remains within the confidence interval. The realism evaluation should therefore be read as evidence that no gross discrepancy was detected, rather than as positive evidence of distributional equivalence. The limited sample size and statistical power of the observer study should therefore be considered when interpreting these results.

\begin{table}[h!]
\centering
\begin{adjustbox}{max width=\linewidth}
\begin{tabular}{ccc}
\hline
Reader&Percent accuracy&Median confidence level\\
\hline
\hline
NM physician 1& 55\% &2\\
\hline
NM physician 2& 45\% &3\\
\hline
NM physician 3& 50\% &3\\
\hline
NM physician 4& 65\% &2\\
\hline
NM physician 5& 55\% &4\\
\hline
Avg.& 54\% &2.8\\
\end{tabular}
\end{adjustbox}
\caption{Performance in the subjective realism evaluation}
\label{tab:Realism_performance}
\end{table}

\subsection{Performance in segmenting inflammation}

\begin{figure}[htbp]
\centering
\includegraphics[width=\linewidth]{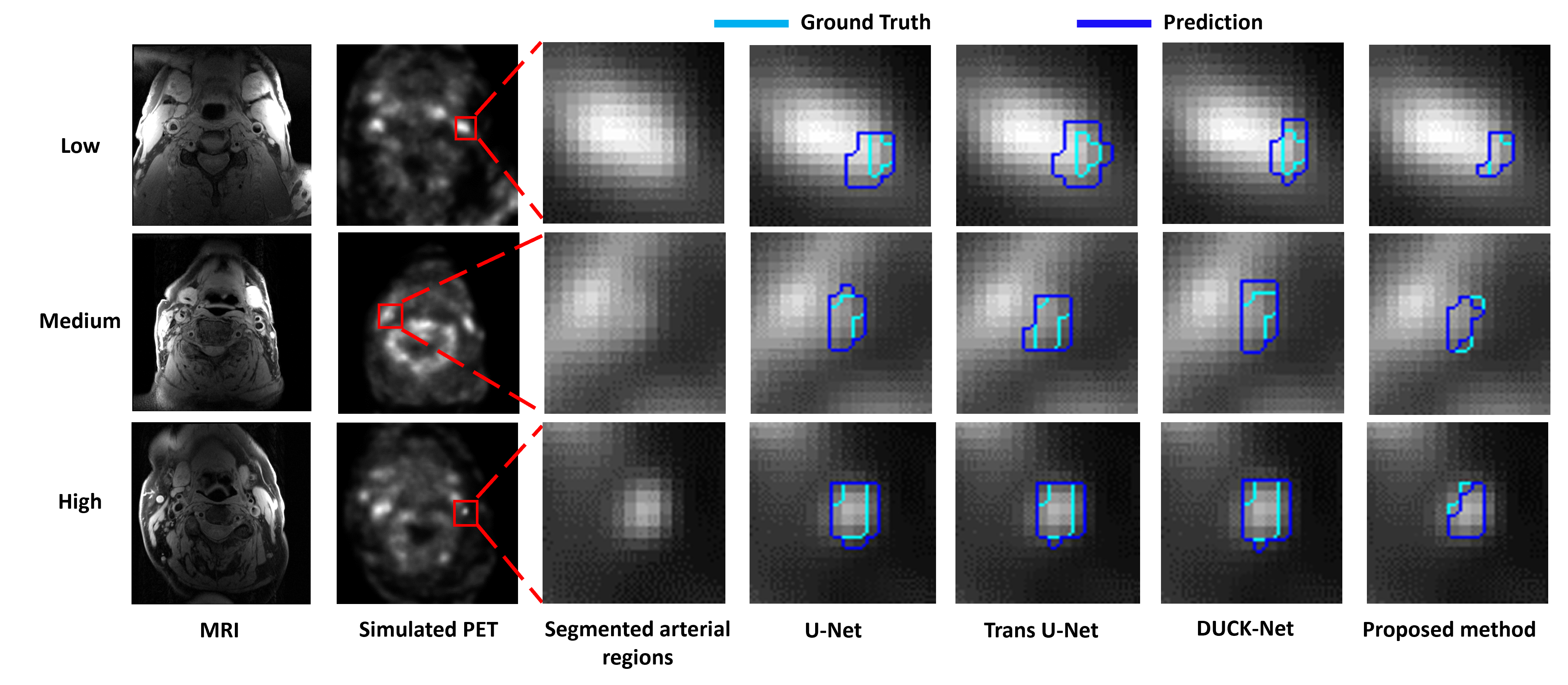}
\caption{Three representative results corresponding to different contrast ranges using proposed method and the other three DL algorithms. From top to bottom row corresponding to: Contrast = 0.83(low), 1.27(medium), and 1.74(high), respectively.}
\label{fig_result}
\end{figure}

The segmentation performance of the proposed method and the three baseline approaches is summarized in Table \ref{tab:Seg_performance}. Representative qualitative results across different contrast levels are shown in Fig. \ref{fig_result}. HD is reported in voxels on the original reconstructed PET grid (4.17 mm in-plane).

\begin{table}[htbp]
\centering

\begin{subtable}{0.95\textwidth}
\centering
\caption{Dice similarity coefficient (DSC)}
\begin{adjustbox}{max width=\linewidth}
\begin{tabular}{lcccc}
\hline
Method & Contrast $\in [0.6,1)$ & Contrast $\in [1,1.4)$ & Contrast $\in [1.4,1.8]$ & Weighted Avg. \\
\hline
U-Net 
& 0.62 [0.57, 0.67] 
& 0.74 [0.70, 0.78] 
& 0.79 [0.75, 0.82] 
& 0.73 [0.70, 0.76] \\
Trans U-Net 
& 0.60 [0.55, 0.65] 
& 0.69 [0.65, 0.73] 
& 0.76 [0.72, 0.80] 
& 0.69 [0.66, 0.72] \\
DUCK-Net 
& 0.63 [0.58, 0.68] 
& 0.75 [0.71, 0.79] 
& 0.81 [0.78, 0.84] 
& 0.75 [0.72, 0.78] \\
Proposed 
& \textbf{0.69 [0.64, 0.73]} 
& \textbf{0.78 [0.74, 0.81]} 
& \textbf{0.86 [0.83, 0.89]} 
& \textbf{0.79 [0.76, 0.82]} \\
\hline
\end{tabular}
\end{adjustbox}
\end{subtable}

\vspace{0.4cm}

\begin{subtable}{0.95\textwidth}
\centering
\caption{Hausdorff distance (HD, voxels)}
\begin{adjustbox}{max width=\linewidth}
\begin{tabular}{lcccc}
\hline
Method & Contrast $\in [0.6,1)$ & Contrast $\in [1,1.4)$ & Contrast $\in [1.4,1.8]$ & Weighted Avg. \\
\hline
U-Net 
& 2.38 [2.20, 2.56] 
& 2.36 [2.19, 2.53] 
& 2.10 [1.95, 2.25] 
& 2.29 [2.18, 2.40] \\
Trans U-Net 
& 2.37 [2.20, 2.54] 
& 2.38 [2.21, 2.55] 
& 2.16 [2.00, 2.32] 
& 2.32 [2.21, 2.43] \\
DUCK-Net 
& 2.09 [1.93, 2.25] 
& 2.29 [2.12, 2.46] 
& 2.19 [2.03, 2.35] 
& 2.23 [2.12, 2.34] \\
Proposed 
& \textbf{1.95 [1.80, 2.10]} 
& \textbf{1.83 [1.69, 1.97]} 
& \textbf{1.81 [1.67, 1.95]} 
& \textbf{1.85 [1.75, 1.95]} \\
\hline
\end{tabular}
\end{adjustbox}
\end{subtable}

\caption{Segmentation performance across different levels of inflammation-to-background contrast. Values are reported as mean with 95\% confidence intervals. The weighted average is computed across contrast intervals with weights proportional to the number of test slices in each interval (32, 98 and 54 slices respectively). Weighted averages were calculated from the underlying unrounded values, so they may differ slightly from values recomputed using the rounded subgroup means shown in the table. The weighted DSC values are 0.73, 0.69, 0.75 and 0.79 for U-Net, Transformer U-Net, DUCK-Net and the proposed method, respectively.}
\label{tab:Seg_performance}
\end{table}

Overall, the proposed method outperformed all baseline methods across all contrast ranges in terms of both DSC and HD. Specifically, the proposed method achieves the highest weighted average DSC of 0.79, compared with 0.75, 0.73 and 0.69 for DUCK-Net, U-Net and Transformer U-Net, respectively. In terms of boundary accuracy, the proposed method yields the lowest weighted average HD of 1.85 voxels, outperforming DUCK-Net (2.23 voxels), U-Net (2.29 voxels), and Transformer U-Net (2.32 voxels).

Patient-cluster bootstrap comparisons (10,000 resamples) were used for inferential analysis. Relative to U-Net, Transformer U-Net and DUCK-Net, the proposed method increased DSC by 0.055 [0.021, 0.088], 0.094 [0.052, 0.133] and 0.041 [0.012, 0.071], with Holm-corrected $p$ values of 0.018, 0.009 and 0.031, respectively. The corresponding reductions in HD were 0.44 [0.18, 0.69], 0.47 [0.21, 0.73] and 0.38 [0.12, 0.64] voxels, with Holm-corrected $p$ values of 0.019, 0.012 and 0.031, respectively. These effect sizes are consistent with the rounded weighted means reported in Table~\ref{tab:Seg_performance}; small differences arise because the bootstrap calculations used the underlying unrounded values. Because the test set contains only five independent patients, these inferential results should be interpreted cautiously, with emphasis placed on the effect sizes and confidence intervals rather than on the nominal $p$ values alone.

\subsection{Performance in quantifying inflammation uptake}
The performance of the proposed method in quantifying inflammation uptake, compared to the other three DL methods, is presented in Table \ref{tab:Quant_performance}. The proposed method exhibits lower ENRMSE values across all three ranges of contrast when compared to other three DL methods. Patient-cluster bootstrap analysis was used to account for correlations among slices from the same patient. The proposed method reduced overall ENRMSE relative to U-Net by 0.35 (95\% CI 0.25--0.43), Transformer U-Net by 0.44 (95\% CI 0.32--0.56), and DUCK-Net by 0.35 (95\% CI 0.20--0.50).
\begin{table}[h!]
\centering
\begin{adjustbox}{max width=\linewidth}
\begin{tabular}{lcccc}
\hline
Method & $[0.6,1)$ & $[1,1.4)$ & $[1.4,1.8]$ & Overall \\
\hline
U-Net & 1.00 [0.87,1.14] & 0.52 [0.46,0.59] & 0.34 [0.29,0.40] & 0.59 [0.52,0.67]\\
Transformer U-Net & 1.07 [0.93,1.22] & 0.63 [0.55,0.72] & 0.44 [0.37,0.52] & 0.68 [0.60,0.77]\\
DUCK-Net & 1.03 [0.89,1.18] & 0.50 [0.44,0.57] & 0.33 [0.28,0.39] & 0.59 [0.51,0.67]\\
Proposed method & \textbf{0.33 [0.27,0.40]} & \textbf{0.26 [0.22,0.31]} & \textbf{0.13 [0.10,0.17]} & \textbf{0.24 [0.20,0.29]}\\
\hline
\end{tabular}
\end{adjustbox}
\caption{ENRMSE of the proposed method compared with three DL approaches by inflammation-to-background contrast. Values are reported with 95\% confidence intervals. Overall ENRMSE is calculated by pooling the squared normalized errors across all test slices before taking the root mean square, rather than by averaging subgroup ENRMSE values.}
\label{tab:Quant_performance}
\end{table}

\subsection{Ablation study and design sensitivity}
\label{sec:ablation}

The contribution of the individual MSR components was evaluated by selectively retaining groups of branches while leaving the remaining training procedure unchanged. Table \ref{tab:ablation_branch} summarizes the results. The full MSR configuration achieved the highest DSC (0.79), lowest HD (1.85 voxels), and lowest ENRMSE (0.24). Removing the double-convolution branch changed these values to 0.78, 1.92 voxels, and 0.26, respectively. The double-convolution-only, standard-residual-only, and dilated-residual-only configurations yielded DSC values of 0.73, 0.76, and 0.75 and ENRMSE values of 0.43, 0.35, and 0.36, respectively. These point estimates support complementary contributions from the branch groups; no separate significance test was performed among the ablation variants.

\begin{table}[htbp]
\centering

\begin{adjustbox}{max width=\linewidth}
\begin{tabular}{lcccc}
\hline
Configuration & DSC & HD (voxels) & ENRMSE & Parameters (M)\\
\hline
Double-convolution only & 0.73 & 2.30  & 0.43 & 8.69\\
Standard residual branches only & 0.76 & 2.02  & 0.35 & 14.31\\
Dilated residual branches only & 0.75  & 2.08  & 0.36 & 10.11\\
Full MSR without double-convolution & 0.78  & 1.92 & 0.26 & 16.90\\
Full MSR & \textbf{0.79} & \textbf{1.85 } & \textbf{0.24} & 18.06\\
\hline
\end{tabular}
\end{adjustbox}
\caption{Ablation results for MSR branch composition. DSC and HD are reported as mean; ENRMSE is reported as a point estimate. Parameter counts were obtained from the implemented PyTorch models.}
\label{tab:ablation_branch}
\end{table}

Addition-based fusion was compared with concatenation after adjusting the concatenation variant to approximately match the total number of model parameters (18.06 M for addition and 18.72 M for concatenation). Addition achieved DSC 0.79, HD 1.85 voxels and ENRMSE 0.24, compared with 0.78, 1.91 voxels and 0.26 for concatenation. The difference was modest but consistent, supporting addition primarily as a parameter-efficient mechanism that maintains a shared channel representation rather than as a universally superior fusion operation.

The morphological mask-enlargement factor was swept over $1\times$, $2\times$, $3\times$, $4\times$, $6\times$ and $8\times$ relative to the original mask size. DSC increased from 0.74 at $1\times$ to a maximum of 0.79 at $4\times$ and then decreased to 0.76 at $8\times$; ENRMSE showed the complementary trend, decreasing from 0.31 to 0.24 before increasing to 0.29. These results support the expected trade-off between excluding spilled-out activity with a narrow mask and reintroducing blood-pool/background activity with an excessively broad mask.

\subsection{Sensitivity to prediction binarisation threshold}
\label{sec:sensitivity}

The prediction threshold used only to convert the continuous fractional output to a binary mask for calculation of DSC and HD was swept from 0.3 to 0.7. As shown in Table~\ref{tab:threshold_sensitivity}, a threshold of 0.5 yielded the lowest HD and tied for the highest DSC, indicating a broad optimum around the selected operating point. This threshold does not affect uptake quantification because ENRMSE is calculated directly from the continuous fractional output.

\begin{table}[htbp]
\centering

\begin{adjustbox}{max width=\linewidth}
\begin{tabular}{ccc}
\hline
Threshold & DSC & HD (voxels)\\
\hline
0.3 & 0.75 & 2.05\\
0.4 & 0.78 & 1.92\\
0.5 & \textbf{0.79} & \textbf{1.85}\\
0.6 & \textbf{0.79} & 1.89\\
0.7 & 0.75 & 2.04\\
\hline
\end{tabular}
\end{adjustbox}
\caption{Sensitivity of segmentation metrics to the prediction binarisation threshold. The threshold is applied only when converting the continuous fractional output to a binary mask for DSC and HD evaluation.}
\label{tab:threshold_sensitivity}
\end{table}

\subsection{Comparison with conventional partial volume correction}
\label{sec:pvc}

To compare the proposed tissue-fraction estimator with a conventional partial-volume-correction approach, iterative Yang correction was applied to the same held-out test data. The lesion region for the Yang correction was defined on the attention-masked PET image using a fixed threshold of 50\% of the within-mask slice maximum. The correction used a 4.0 mm FWHM Gaussian point-spread function, 4.17 mm in-plane spacing and 10 iterations. Because iterative Yang produces corrected uptake rather than a tissue-fraction map, the comparison was performed on the downstream task of inflammation-uptake quantification using the same ENRMSE definition as for the deep-learning methods. Table~\ref{tab:yang_comparison} reports the results by inflammation-to-background contrast. Iterative Yang yielded ENRMSE values of 1.03, 0.47 and 0.29 across the three contrast intervals, whereas the proposed method yielded 0.33, 0.26 and 0.13. The corresponding overall ENRMSE values were 0.57 and 0.24, respectively. The largest difference occurred in the low-contrast interval.

\begin{table}[htbp]
\centering

\begin{adjustbox}{max width=\linewidth}
\begin{tabular}{lcccc}
\hline
Method & $[0.6,1.0)$ & $[1.0,1.4)$ & $[1.4,1.8]$ & Overall\\
\hline
Iterative Yang & 1.03 & 0.47 & 0.29 & 0.57\\
Proposed method & \textbf{0.33} & \textbf{0.26} & \textbf{0.13} & \textbf{0.24}\\
\hline
\end{tabular}
\end{adjustbox}
\caption{Comparison of uptake-quantification error between iterative Yang partial-volume correction and the proposed tissue-fraction estimation method. Values are ENRMSE; lower values indicate more accurate uptake quantification. The overall value is calculated from the pooled squared normalized errors across the 32, 98 and 54 test slices in the three contrast intervals, respectively.}
\label{tab:yang_comparison}
\end{table}

\subsection{Quantitative analysis on real patient data}

\begin{figure}[htbp]
\centering
\includegraphics[width=\linewidth]{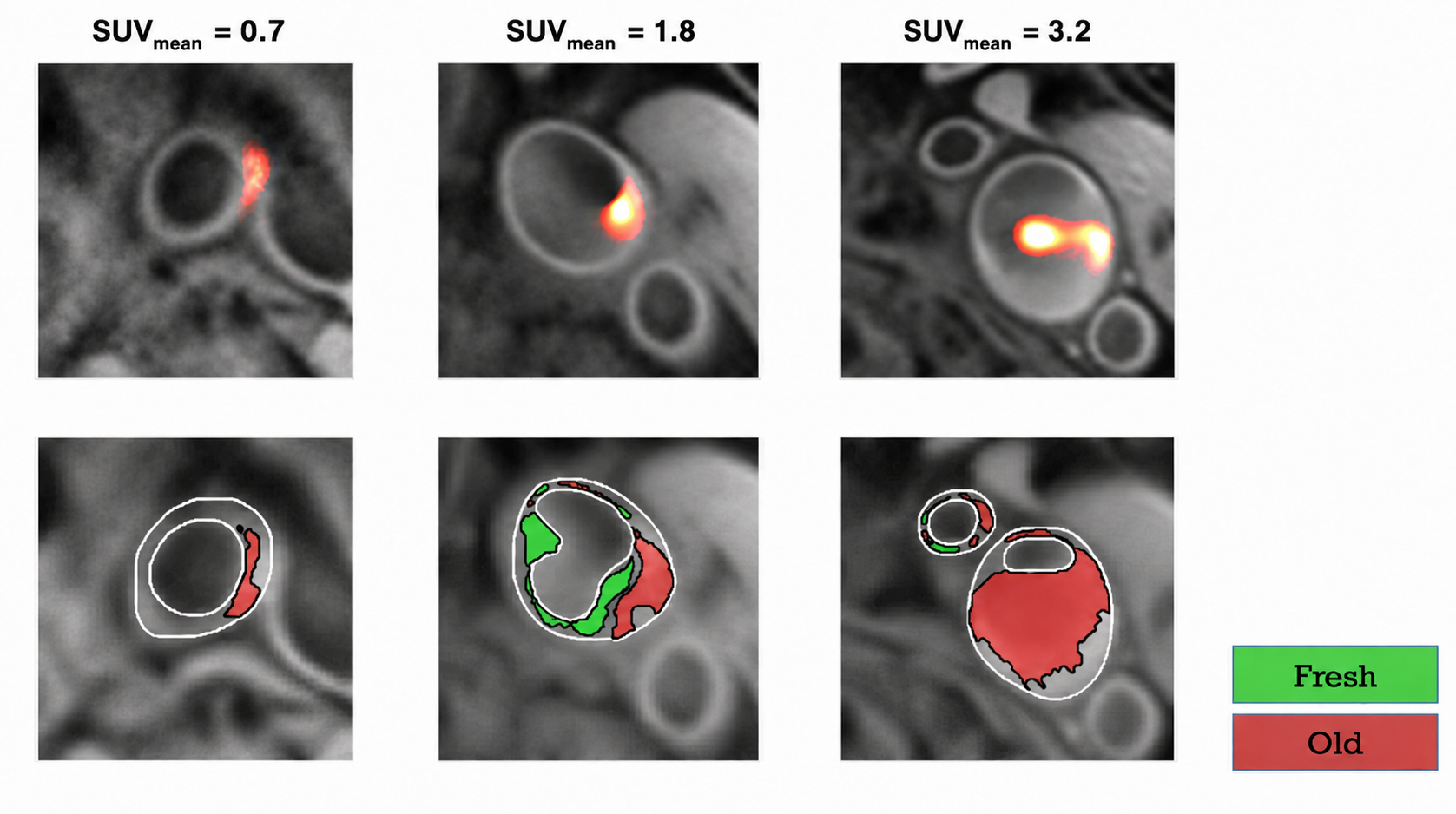}
\caption{Focal radiotracer uptake on PET and corresponding carotid plaque-component segmentation on MR images for three representative cases. The MR images are displayed at the same image scale for direct visual comparison. The MR plaque-component segmentation was generated using the previously published two-stage multi-weighted MRI method of \citet{Intro_two_stage}. In the plaque segmentation, green indicates fresh hemorrhage within 10 days, and red indicates old hemorrhage over 10 days. The mean SUV is measured and shown in the figure.}
\label{fig_real_petmr}
\end{figure}

\begin{figure}[htbp]
\centering
\includegraphics[width=\linewidth]{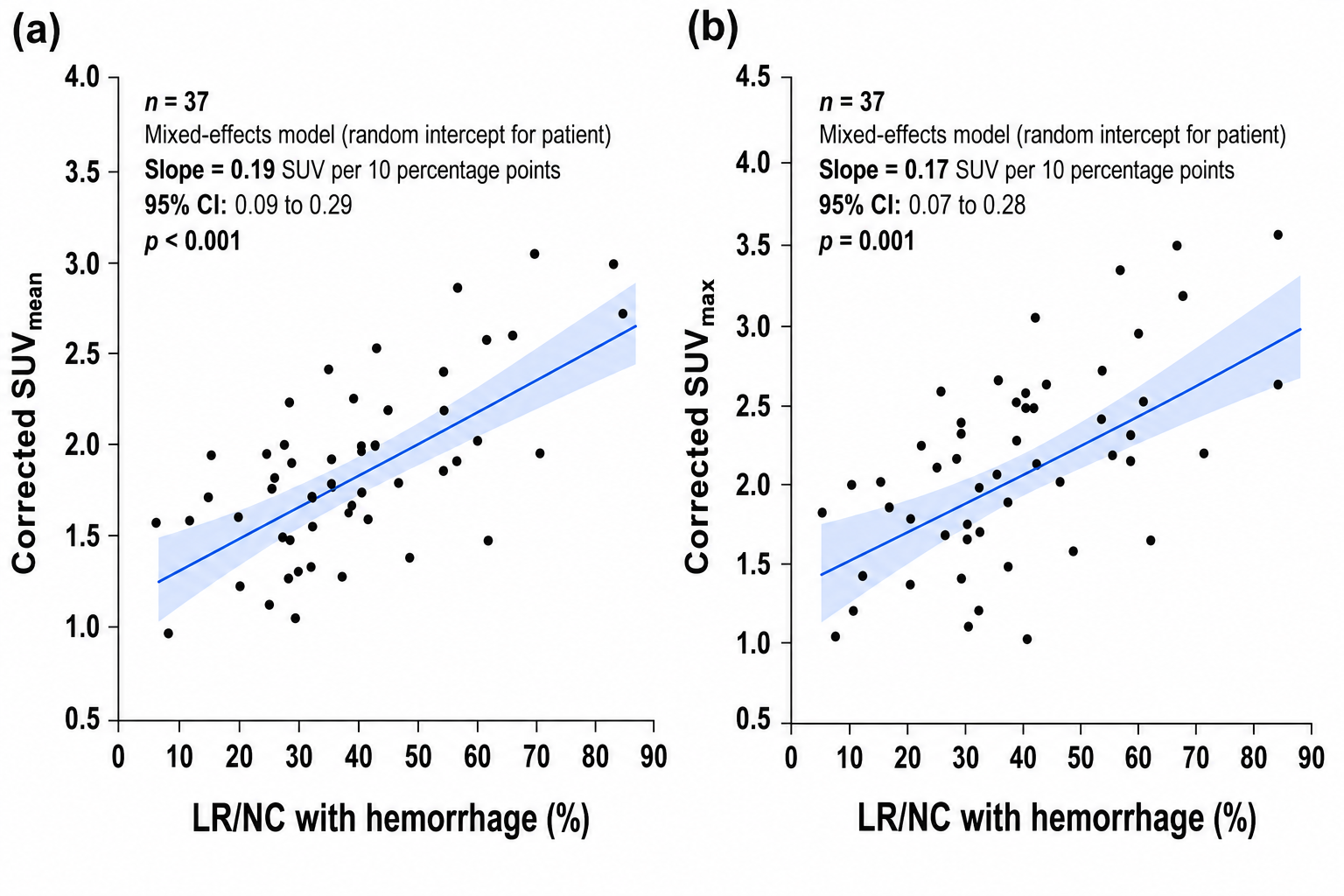}
\caption{Association between the percentage volume of LR/NC with hemorrhage and corrected $SUV_{mean}$ (a) and corrected $SUV_{max}$ (b). Points represent slice-level observations. Solid lines indicate population-level fixed-effect fits from linear mixed-effects models with patient-specific random intercepts, and shaded bands indicate 95\% confidence intervals for the fixed-effect mean.}
\label{fig_real_result}
\end{figure}

Figure~\ref{fig_real_petmr} shows the spatial correspondence between regions of increased uptake on PET images and the appearance of hemorrhage on MR images. Regions with elevated radiotracer uptake on PET are observed to co-localize with areas identified as hemorrhage on MR, providing qualitative evidence of the association between inflammation and plaque composition.
Figure~\ref{fig_real_result} illustrates the association between the percentage volume of LR/NC and the corrected $SUV_{mean}$ and $SUV_{max}$. The analysis included 37 slice-level observations from seven patients and used linear mixed-effects models with a patient-specific random intercept. For corrected $SUV_{mean}$, a 10-percentage-point increase in LR/NC with hemorrhage was associated with a 0.19-unit increase in SUV (95\% CI 0.09--0.29; $p<0.001$). For corrected $SUV_{max}$, the corresponding slope was 0.17 (95\% CI 0.07--0.28; $p=0.001$). These findings indicate moderate positive imaging associations, but given the small number of patients they should be considered exploratory. A Bland--Altman analysis was not performed because LR/NC percentage volume and corrected SUV are different quantities in different units rather than alternative measurements of a common measurand. Their relationship is therefore appropriately framed as association rather than agreement.

\section{Discussion}

Reliable detection of inflammation within the carotid wall poses challenges due to PVEs and spillover effects, impeding studies in this domain. Towards addressing this challenge, we propose a DL-based PET segmentation approach for accurate inflammation segmentation and uptake quantification. Utilizing a customized $^{64}$Cu-CANF-comb PET simulation for training, an approach that is designed to estimate tissue-fraction effects, and a two-stage U-Net structure with a customized MSR backbone, our methodology yields promising results in segmentation and quantification tasks compared to alternative DL-based methods.

To assess the realism of the synthetic PET images, we conducted a 2AFC study with expert physician readers. The readers correctly identified the real PET image approximately 50\% of the time, indicating that they could not reliably discriminate simulated from real PET images under the tested conditions. This result supports the perceptual realism of the simulated images within the scope of the observer study.

The proposed method uses the strategy of estimating the ground truth of inflammation as a fractional volume instead of a binary mask. This offers the ability to account for PVEs arising due to tissue-fraction effects, enabling not only accurate segmentation of inflammation but also reliable measurement of inflammation uptake under the influence of PVEs. The quantification results presented in Table~\ref{tab:Quant_performance} highlight the improvement in accuracy of our proposed method in estimating inflammation uptake. Lower ENRMSE values obtained with the MSR backbone illustrate more reliable quantification compared to the other three DL-based methods under consideration. This suggests the improved performance of the proposed method in quantifying inflammation uptake across different contrast levels.

In this study, the proposed method, incorporating the customized MSR backbone, demonstrates superior segmentation performance compared to conventional automated segmentation approaches. The improved performance is reflected by higher DSCs and lower HDs across different contrast levels, indicating enhanced accuracy and robustness in segmenting regions of inflammation. These results validate the effectiveness of the MSR backbone in capturing multi-scale features and improving segmentation quality. The advantages of the proposed method can be attributed to the use of convolution-based design, which benefits from inherent properties such as shift invariance and enables stable learning under limited training data conditions. Leveraging this property, the MSR backbone is designed to capture both global contextual information and fine-grained features, contributing to improved segmentation performance. Notably, the Transformer-based U-Net, which has demonstrated strong performance in other medical imaging tasks, showed relatively poorer performance in this study. One possible explanation is the limited dataset size and the absence of external pretraining, as Transformer-based models can be more data demanding when trained from scratch \citep{Liu2021SmallVT,Lu2022SmallVT}. This explanation was not directly tested in the present study.

 The improved performance of the proposed MSR backbone is consistent with the contribution of the multi-scale convolutional blocks, as supported by the branch and fusion ablation results. This design not only enables our method to learn long-range relations of different signals but also extract finer detailed features, thus enhancing segmentation performance. The representative segmentation results shown in Fig.~\ref{fig_result} provide visual evidence of the effectiveness of our method in accurately segmenting regions of inflammation across different contrast ranges. Even in low contrast ranges, while our method may not segment inflammation accurately, it can still localize the regions of inflammation. These examples illustrate the ability of our approach to handle variations in contrast levels and produce reliable segmentation results.

The pilot clinical quantitative analysis presented in this study provides insights into the relationship between inflammation and LR/NC within the carotid arteries. Our findings reveal a positive association between the percentage volume of LR/NC with hemorrhage and PET-derived inflammation SUVs. This association suggests a potential link between the extent of lipid-rich or necrotic cores and inflammation uptake within the carotid arteries. The observed positive association is consistent with existing literature relating inflammation to advanced plaque features. Because the analysis included only seven patients and lacked histopathological ground truth, this result should be considered preliminary and hypothesis-generating rather than evidence of clinical utility. It should be noted that completely accurate expert annotations are not available, as PET images inherently lack a definitive ground truth for inflammation. Future work may involve validation using alternative approaches, such as histopathology, to further confirm the clinical relevance of these findings.

Despite the advancements of the proposed method, there are several limitations to consider in this study. Firstly, the proposed simulation pipeline has its constraints. Achieving high-quality simulation requires the utilization of the GATE simulation toolkit, which models major physical processes during PET scanning. However, this simulation is time-consuming and requires significant computing resources. The entire simulation process for a single patient takes at least one week, thereby limiting the availability of our proposed method and the size of the simulation dataset. We expect that enhancements to the GATE simulation toolkit, such as support for parallel computing on GPUs, could accelerate the simulation process and address this issue. Secondly, the range of radioactivity of synthetic inflammation is based on statistical measurements and empirical evaluations conducted on a limited number of patient data. Although the range applied in this study is derived from inflammations delineated by several experienced radiologists, it still limits the generalizability of this study due to the constraints of the patient database size.

Additional limitations include simulation-to-reality domain shift, the single-center design, observer variability in manual vessel-wall delineation and PET--MR alignment, and the absence of voxel-wise uncertainty estimates. Although some readers detected subtle differences between simulated and real PET images, the source and potential impact of these differences were not evaluated in this study. Therefore, the observer study does not establish distributional equivalence or guarantee performance on real PET data. Inter-operator registration reproducibility was not prospectively measured. Because $^{64}$Cu-CANF-comb is an investigational tracer with limited clinical availability, an independent external cohort with comparable PET/MR data was not available. No histopathological reference was available; therefore, the patient analysis establishes an imaging association rather than biological validation of inflammation. The clinical association analysis included repeated slice-level observations from only seven patients; although patient-specific random intercepts were used to account for within-patient dependence, the small number of independent patients limits precision and generalizability, and the results should therefore be regarded as exploratory. These factors may cause performance measured on simulated data to overestimate performance on clinical data. Future studies should recalibrate the simulation parameters using larger and multi-center clinical datasets and validate, and where appropriate fine-tune, the model on independent real PET data.

Finally, the evaluation of the proposed method utilized figures of merit including DSC and HD, but the performance of these metrics may not directly correspond to clinical performance~\citep{Discussion_DSC_HD}. To ensure the accuracy, reliability, and generalizability of the simulation-based study, rigorous validation in real-world clinical settings is essential. While the pilot clinical evaluation conducted in this study yielded promising results, further work is necessary to validate the proposed method using a larger cohort of real patient data, ideally paired with corresponding histopathological information. This validation process will serve to further confirm the accuracy and robustness of the proposed approach across diverse patient populations and imaging conditions. Additionally, efforts toward assessing the generalizability of the proposed method by evaluating its performance on datasets from different institutions and imaging protocols are important future research directions. Validating the proposed approach in real clinical scenarios will help to establish its credibility and applicability for widespread use in medical practice.

\section{Conclusion}
Our study presents an advanced DL-based method for inflammation segmentation within the carotid wall. By utilizing a customized PET simulation pipeline and an MSR backbone, the proposed methodology demonstrates improved segmentation accuracy and quantification precision compared to three alternative methods under simulated conditions. According to the RELAINCE guidelines~\citep{Conclusion}, the supported claim is: ``A DL-based image segmentation method for quantifying uptake within carotid plaque achieved lower ENRMSE than three comparator DL-based methods in clinically realistic simulation studies of $^{64}$Cu-CANF-comb carotid PET.'' The patient analysis included seven patients and should be considered preliminary. Larger multi-center studies and validation against histopathology are required before clinical deployment.

\section*{Ethics statement}
This study was conducted in accordance with the Declaration of Helsinki and was approved by the Institutional Review Board of Washington University in St.\ Louis. Written informed consent was obtained from all participants prior to imaging.

\section*{Data availability statement}
The patient imaging data that support the findings of this study cannot be made publicly available because they contain potentially identifiable clinical information. De-identified data supporting the findings may be available from the corresponding author upon reasonable request and subject to institutional and regulatory approval.

\section*{Conflict of interest}
All authors declare that they have no known competing financial interests or personal relationships that could have appeared to influence the work reported in this paper.

\section*{Acknowledgments}
This work was supported in part by the National Institutes of Health (NIH) under grant numbers R01HL132600, R01HL150891, R01EB031051, and R56EB028287.

\appendix
\section{Receptive field and approximation analysis for the MSR block}
\label{app:theory}
\setcounter{equation}{0}
\renewcommand{\theequation}{A.\arabic{equation}}

\emph{Receptive field.} For a sequence of convolutional layers with kernel sizes $k_l$, strides $s_l$ and dilations $d_l$ $(l=1,\dots,L)$, the effective kernel size of layer $l$ is $k_l^{\mathrm{eff}} = k_l + (k_l-1)(d_l-1)$, and the receptive field of the output with respect to the input is
\begin{equation}
    \mathrm{RF} = 1 + \sum_{l=1}^{L} \left(k_l^{\mathrm{eff}} - 1\right)\prod_{j<l} s_j ,
\end{equation}
where $\prod_{j<l}s_j$ accounts for the spatial scaling introduced by all earlier layers. A single-scale backbone realises one value of $\mathrm{RF}$; the MSR block realises $M$ values $\{\mathrm{RF}_1,\dots,\mathrm{RF}_M\}$ in parallel.

\emph{Fusion.} Let $B_i$ denote the mapping realised by branch $i$. Fusion is performed by pixel-wise addition,
\begin{equation}
    \hat{F}(x) = \sum_{i=1}^{M} B_i(x),
    \label{eq:fusion}
\end{equation}
which combines detail-preserving small-RF branches with context-aware large-RF branches without expanding the channel dimension.

\emph{Approximation bound.} Let the target admit a decomposition into scale-localised components, $f(x) = \sum_{i=1}^{M} f_i(x)$. With the fusion of Equation~(\ref{eq:fusion}) the estimate is $\hat{f} = \sum_i B_i$, and by the triangle inequality
\begin{equation}
    \left\| f - \hat{f} \right\|_{L^2(\Omega)}
    = \left\| \sum_{i=1}^{M} (f_i - B_i) \right\|_{L^2(\Omega)}
    \le \sum_{i=1}^{M} \left\| f_i - B_i \right\|_{L^2(\Omega)} .
\end{equation}
Each branch therefore need only approximate the component matched to its effective RF. If the residuals $f_i - B_i$ are additionally taken to be weakly correlated across branches, the bound tightens to
\begin{equation}
    \left\| f - \hat{f} \right\|_{L^2(\Omega)}
    \le \sqrt{\sum_{i=1}^{M} \left\| f_i - B_i \right\|_{L^2(\Omega)}^2 } .
\end{equation}

We note two limitations of this analysis, which are the reason the design is justified empirically in Section~\ref{sec:ablation} rather than analytically. First, the bounds are upper bounds and do not establish that the MSR decomposition is tighter than that achieved by a single-scale network of equal capacity. Second, the weak-correlation assumption underlying Equation~(A.4) is not verified for PET data; branches operating on the same input may in practice produce correlated residuals, in which case only the looser bound of Equation~(A.3) applies.

\bibliographystyle{abbrvnat}
\bibliography{references}

\end{document}